\documentclass[pra,aps,twocolumn,showpacs]{revtex4}
\usepackage{amsmath,amsfonts,amssymb,graphics,graphicx,epsfig}
\usepackage[latin1]{inputenc}
\usepackage[usenames,dvipsnames]{color}

\begin{document}

\title{Short-time vs. long-time dynamics of entanglement in quantum lattice
models}
\author{R. G. Unanyan}
\email{unanyan@physik.uni-kl.de}
\affiliation{Department of Physics and Research Center OPTIMAS, University of
Kaiserslautern, D-67663 Kaiserslautern, Germany}
\author{D. Muth}
\affiliation{Department of Physics and Research Center OPTIMAS, University of
Kaiserslautern, D-67663 Kaiserslautern, Germany}
\author{M. Fleischhauer}
\affiliation{Department of Physics and Research Center OPTIMAS, University of
Kaiserslautern, D-67663 Kaiserslautern, Germany}
\date{\today}

\begin{abstract}
We study the short-time evolution of the bipartite entanglement in quantum
lattice systems with local interactions in terms of the purity of the
reduced density matrix. A lower bound for the purity is derived in terms of
the eigenvalue spread of the interaction Hamiltonian between the partitions.
Starting from an initially separable state the purity decreases as $1 -
(t/\tau)^2$, i.e. quadratically in time, with a characteristic time scale $%
\tau$ that is inversely proportional to the boundary size of the subsystem,
i.e., as an area--law. For larger times an exponential lower bound is
derived corresponding to the well-known linear-in-time bound of the
entanglement entropy. The validity of the derived lower bound is illustrated
by comparison to the exact dynamics of a 1D spin lattice system as well as a
pair of coupled spin ladders obtained from numerical simulations.
\end{abstract}

\maketitle





\section*{I -- Introduction}


Motivated by the question whether the time evolution of interacting quantum
systems can be efficiently simulated with the help of matrix-product
decompositions of the many-body wave function \cite{Verstraete2008,Vidal2003,Vidal2004} or corresponding analogues in higher dimensions \cite%
{Nishino1998,Verstraete2004c,Murg2007,Vidal2007a,Verstraete2008,Sierra1998}, the dynamics of
entanglement in quantum lattice models has become an important research area
in quantum physics. A convenient measure of entanglement are \cite%
{Schuch2008a} the von Neumann and R\'enyi entropies. It was shown in \cite%
{Bravyi2006} and \cite{Eisert2006} that the von Neumann entropy of a
subsystem which starts in an initially separable state has an upper bound
that grows linear in time. The linear growth of the entropy, as observed by
Calabrese and co-workers \cite{Calabrese2005,Fagotti2008}, which corresponds to an
exponential growth of the effective bond dimension of matrix--product states
(MPS) represents a severe limitation for the simulability of the unitary
time evolution of quantum many-body systems. In the present note we derive
an upper bound to the bipartite entanglement that also holds for short
times. In particular we consider the purity of the reduced density matrix of
one of the partitions. General quantum mechanical arguments suggest that the
purity cannot decease exponentially for short times as implied by the linear
entanglement bound \cite{Lieb1972,Bravyi2006,Eisert2006}, but rather
quadratically. Here we derive a quadratic lower bound for the purity. This
finding which is the main result of our paper has practical relevance for
the numerical simulation of another class of dynamical problems that gained
a lot of interest recently where the time evolution is non-unitary due to a
coupling to external reservoirs \cite%
{Syassen2008,Durr2009,Garcia-Ripoll2009,Prosen2008,Prosen2008a,Prosen2009,Diehl2008,Daley2009,Han2009,Kantian2009,Kiffner2009}%
. The non-unitary Liouvillian dynamics of the system density matrix is
equivalent to a time evolution of the many-body wave function with a complex
Hamiltonian and with a stochastic sequence of projections, called quantum
jumps \cite{Carmichael1993,Dalibard1992,Molmer1993}. If the frequency of
such projections is sufficiently large they can prevent the growth of
entanglement within the system by a mechanism similar to the well-known
quantum Zeno effect \cite{Facchio2001a}. As a result of this the time
evolution of open system may be simulated using an adaptive MPS expansion as
e.g. within the time evolving block decimation algorithm (TEBD) \cite%
{Vidal2003,Vidal2004} for longer times. The critical frequency of such a
Zeno effect for entanglement is determined by the coefficient of the
quadratic term in the short time expansion of the purity. Note that this
effect would be absent for an exponential time dependence of the purity. For
larger times we derive an exponential lower bound corresponding to the
well-known linear--in--time bound of the entanglement entropy \cite%
{Lieb1972,Bravyi2006,Eisert2006}. Although this result follows directly from
the latter entropy bound, we derive it here in a few lines without making
use of the somewhat more involved proof of the linear entropy scaling. To
illustrate the validity of our findings we discuss as an example the 1D spin-%
$\frac{1}{2}$ XX and XXZ models, where we calculate the time evolution of
the purity using the numerical time-evolving block decimation algorithm \cite%
{Vidal2003,Vidal2004}, as well as two coupled spin chains, which allow us to
illustrate the scaling with the boundary size.


\section*{II -- Short-time behavior}


We here consider a lattice model with a bipartition into parts $A$ and $B$,
which are, say, compact sets of lattice sites. We assume that the
Hamiltonian of the total systems can be written as
\begin{equation}
\hat{H}_{AB}=\hat{H}_{A}+\hat{H}_{B}+\sum_{q\in \{A|B\}}\hat{H}_{q}^{\left(
A\right) }\otimes \hat{H}_{q}^{\left( B\right) },  \label{eq:hamiltonian}
\end{equation}%
where $\hat{H}_{A}$ and $\hat{H}_{B}$ are the parts of the Hamiltonian
acting on lattice sites inside the respective partitions. The last sum
describes the interaction between the two parts an extends over all bonds,
labeled by the index $q$, that connect sites from both partitions. We assume
an interaction that has strict finite range. In this case the total number
of bonds scales with the size $N$ of the surface separating the two
partitions.

Any pure state of the total system can be decomposed as
\begin{equation}
\left\vert \Psi \left( t\right) \right\rangle _{AB}={\displaystyle%
\sum\limits_{\alpha =1}^{L}}\sqrt{\xi _{\alpha }}\left\vert \phi _{\alpha
}^{\left( A\right) }\right\rangle \otimes \left\vert \phi _{\alpha }^{\left(
B\right) }\right\rangle ,  \label{eq:Schmidt}
\end{equation}%
where $\left\vert \phi _{\alpha }^{\left( A\right) }\right\rangle $ and $%
\left\vert \phi _{\alpha }^{\left( B\right) }\right\rangle $ are orthonormal
sets of states of the subsystems, $\xi _{\alpha }\geq 0$ are the Schmidt
coefficients, and $L$ is at most the dimension of the Hilbert space of the
smaller subsystem. As $\left\vert \Psi \left( t\right) \right\rangle _{AB}$
is normalized, $\sum\limits_{\alpha }\xi _{\alpha }=1$. The reduced density
operator of the subsystem $A$, $\rho _{A}=\mathrm{tr}_{B}\{\rho _{AB}\}$,
where $\rho _{AB}=\left\vert \Psi \left( t\right) \right\rangle
_{AB}\left\langle \Psi \left( t\right) \right\vert _{AB}$, satisfies the
equation of motion
\begin{equation}
\frac{\mathrm{d}\rho _{A}}{\mathrm{d}t}=-i\,\mathrm{tr}_{B}\Bigl\{\left[
\hat{H}_{AB},\rho _{AB}\right] \Bigr\}.  \label{eq:densityEquation}
\end{equation}%
(Note that we have set $\hbar=1$ throughout this text.) The form of the Hamiltonian (\ref{eq:hamiltonian}) and the cyclic property
of the trace allows us to obtain the following equation
\begin{equation}
\frac{\mathrm{d}}{\mathrm{d}t}\mathrm{tr}\rho _{A}^{2}=-2i\mathrm{tr}\left\{
\rho _{A}\mathrm{tr}_{B}\left[ \sum_{q}\hat{H}_{q}^{\left( A\right) }\otimes
\hat{H}_{q}^{\left( B\right) },\rho _{AB}\right] \right\}
\label{purityrate1}
\end{equation}%
for the purity rate. The traces on the right hand side can be calculated in the
eigenbasis of the $\rho _{A}$ and $\rho _{B}$:
\begin{widetext}
\begin{equation}
\mathrm{tr}\left\{ \rho _{A}\mathrm{tr}_{B}\left( \sum_{q}\hat{H}%
_{q}^{\left( A\right) }\otimes \hat{H}_{q}^{\left( B\right) }\cdot \rho
_{AB}\right) \right\} =\sum\limits_{q,\alpha }\xi _{\alpha }\left\langle
\phi _{\alpha }^{\left( A\right) }\right\vert \mathrm{tr}_{B}\hat{H}%
_{q}^{\left( A\right) }\otimes \hat{H}_{q}^{\left( B\right) }\rho
_{AB}\left\vert \phi _{\alpha }^{\left( A\right) }\right\rangle =
\label{firstterm}
\end{equation}%
\begin{equation*}
\sum\limits_{q,\alpha ,\alpha ^{\prime }}\xi _{\alpha }\sqrt{\xi _{\alpha
}\xi _{\alpha ^{\prime }}}\left\langle \phi _{\alpha }^{\left( A\right)
}\right\vert \mathrm{tr}_{B}\hat{H}_{q}^{\left( A\right) }\otimes \hat{H}%
_{q}^{\left( B\right) }\left\vert \phi _{\alpha ^{\prime }}^{\left( A\right)
}\right\rangle \otimes \left\vert \phi _{\alpha ^{\prime }}^{\left( B\right)
}\right\rangle \left\langle \phi _{\alpha }^{\left( B\right) }\right\vert =
\end{equation*}%
\begin{equation*}
=\sum\limits_{q,\alpha ,\alpha ^{\prime }}\xi _{\alpha }\sqrt{\xi _{\alpha
}\xi _{\alpha ^{\prime }}}\left\langle \phi _{\alpha }^{\left( A\right)
}\right\vert \hat{H}_{q}^{\left( A\right) }\left\vert \phi _{\alpha ^{\prime
}}^{\left( A\right) }\right\rangle \mathrm{tr}_{B}\left( \hat{H}_{q}^{\left(
B\right) }\left\vert \phi _{\alpha ^{\prime }}^{\left( B\right)
}\right\rangle \left\langle \phi _{\alpha }^{\left( B\right) }\right\vert
\right) =
\end{equation*}%
\begin{equation}
=\sum\limits_{q,\alpha ,\alpha ^{\prime }}\xi _{\alpha }\sqrt{\xi _{\alpha
}\xi _{\alpha ^{\prime }}}\left\langle \phi _{\alpha }^{\left( A\right)
}\right\vert \hat{H}_{q}^{\left( A\right) }\left\vert \phi _{\alpha ^{\prime
}}^{\left( A\right) }\right\rangle \left\langle \phi _{\alpha }^{\left(
B\right) }\right\vert \hat{H}_{q}^{\left( B\right) }\left\vert \phi _{\alpha
^{\prime }}^{\left( B\right) }\right\rangle ,  \label{term1}
\end{equation}%
and in the same way, the second term of the commutator is
\begin{equation}
\mathrm{tr}\left\{ \rho _{A}\mathrm{tr}_{B}\left( \rho _{AB}\cdot \sum_{q}%
\hat{H}_{q}^{\left( A\right) }\otimes \hat{H}_{q}^{\left( B\right) }\right)
\right\} =\sum\limits_{q,\alpha ,\alpha ^{\prime \prime }}\xi _{\alpha }%
\sqrt{\xi _{\alpha }\xi _{\alpha ^{\prime \prime }}}\left\langle \phi
_{\alpha ^{\prime \prime }}^{\left( A\right) }\right\vert \hat{H}%
_{q}^{\left( A\right) }\left\vert \phi _{\alpha }^{\left( A\right)
}\right\rangle \left\langle \phi _{\alpha ^{\prime \prime }}^{\left(
B\right) }\right\vert \hat{H}_{q}^{\left( B\right) }\left\vert \phi _{\alpha
}^{\left( B\right) }\right\rangle .  \label{secondterm}
\end{equation}
\end{widetext}
Combining eqs. (\ref{purityrate1}), (\ref{term1}) and (\ref{secondterm}) one
obtains the following differential equation for the purity
\begin{eqnarray}
&&\frac{\mathrm{d}}{\mathrm{d}t}\mathrm{tr}\rho
_{A}^{2}=-2i\sum\limits_{q,\alpha ,\beta }\sqrt{\xi _{\alpha }\xi _{\beta }}%
\left( \xi _{\alpha }-\xi _{\beta }\right) \times  \label{eq:purity} \\
&&\quad \times \left\langle \phi _{\alpha }^{\left( A\right) }\right\vert
\hat{H}_{q}^{\left( A\right) }\left\vert \phi _{\beta }^{\left( A\right)
}\right\rangle \left\langle \phi _{\alpha }^{\left( B\right) }\right\vert
\hat{H}_{q}^{\left( B\right) }\left\vert \phi _{\beta }^{\left( B\right)
}\right\rangle .  \notag
\end{eqnarray}%
This equation can be rewritten in a compact form
\begin{equation}
\frac{\mathrm{d}}{\mathrm{d}t}\mathrm{tr}\rho _{A}^{2}=\mathrm{tr}\left[
\Theta \cdot Q\right] ,  \label{eq:purityrate}
\end{equation}%
where
\begin{equation}
\Theta _{\alpha \beta }=-2i\sqrt{\xi _{\alpha }\xi _{\beta }}\left( \xi
_{\alpha }-\xi _{\beta }\right)
\end{equation}%
and
\begin{equation}
Q_{\alpha \beta }=\sum\limits_{q}\left\langle \phi _{\alpha }^{\left(
A\right) }\right\vert \hat{H}_{q}^{\left( A\right) }\left\vert \phi _{\beta
}^{\left( A\right) }\right\rangle \left\langle \phi _{\alpha }^{\left(
B\right) }\right\vert \hat{H}_{q}^{\left( B\right) }\left\vert \phi _{\beta
}^{\left( B\right) }\right\rangle
\end{equation}%
The matrix $\Theta $ has only two non-zero eigenvalues. Indeed, $\Theta $
can be written as
\begin{equation}
\Theta =-2i\left\vert a\right\rangle \left\langle b\right\vert +2i\left\vert
b\right\rangle \left\langle a\right\vert ,  \label{eq:Teta}
\end{equation}%
where
\begin{align*}
\left\vert a\right\rangle & =\left( \xi _{1}^{3/2},\xi _{2}^{3/2},\dots \xi
_{L}^{3/2}\right) ^{T}, \\
\left\vert b\right\rangle & =\left( \xi _{1}^{1/2},\xi _{2}^{1/2},\dots \xi
_{L}^{1/2}\right) ^{T}.
\end{align*}%
It is easy to show that the nonzero eigenvalues of $\Theta $ are
\begin{eqnarray}
\lambda _{\pm }\left( \Theta \right) &=&\pm 2\sqrt{\left\langle a\right.
\left\vert a\right\rangle \left\langle b\right\vert \left. b\right\rangle
-\left\vert \left\langle a\right\vert \left. b\right\rangle \right\vert ^{2}}
\label{eq:spectrum} \\
&=&\pm 2\sqrt{\mathrm{tr}\rho _{A}^{3}-\left( \mathrm{tr}\rho
_{A}^{2}\right) ^{2}}.  \notag
\end{eqnarray}%
Let $\left\vert q_{\pm }\right\rangle $ be the corresponding eigenvectors of
$\Theta $, then the trace (\ref{eq:purityrate}) can be evaluated in the
eigenbasis of $\Theta $ which yields the following equation for the purity
rate
\begin{equation}
\frac{\mathrm{d}}{\mathrm{d}t}\mathrm{tr}\rho _{A}^{2}=2\sqrt{\mathrm{tr}%
\rho _{A}^{3}-\left( \mathrm{tr}\rho _{A}^{2}\right) ^{2}}\Bigl(\left\langle
q_{+}\right\vert Q\left\vert q_{+}\right\rangle -\left\langle
q_{-}\right\vert Q\left\vert q_{-}\right\rangle \Bigr).  \label{eq:inequ}
\end{equation}%
The right side of this equation can be bounded from above by the spread of
the eigenvalues of the interaction Hamiltonian between partitions $A$ and $B$%
:
\begin{equation}
\frac{\mathrm{d}}{\mathrm{d}t}\mathrm{tr}\rho _{A}^{2}\leq 2\sqrt{\mathrm{tr}%
\rho _{A}^{3}-\left( \mathrm{tr}\rho _{A}^{2}\right) ^{2}}\Bigl[\lambda
_{\max }-\lambda _{\min }\Bigr]  \label{eq:upperbound}
\end{equation}%
Here $\lambda _{\max }$ and $\lambda _{\min }$ are the maximum and minimum
eigenvalues of $\sum_{q\in \{A|B\}}\hat{H}_{q}^{(A)}\otimes \hat{H}%
_{q}^{(B)} $.

In a similar way one can show that
\begin{equation}
\frac{\mathrm{d}}{\mathrm{d}t}\mathrm{tr}\rho_{A}^{2}\geq-2\sqrt{\mathrm{tr}%
\rho_{A}^{3}-\left(\mathrm{tr}\rho_{A}^{2}\right) ^{2}}\Bigl[ \lambda_{\max}
-\lambda_{\min} \Bigr] .  \label{eq:importinequality}
\end{equation}
Combining the inequalities (\ref{eq:upperbound}) and (\ref%
{eq:importinequality}) we obtain
\begin{equation}
\left\vert \frac{\mathrm{d}}{\mathrm{d}t}\mathrm{tr}\rho_{A}^{2}\right\vert
\leq2\mu\sqrt{\mathrm{tr}\rho_{A}^{3}-\left( \mathrm{tr}\rho_{A}^{2}\right)
^{2}},  \label{final0}
\end{equation}
where
\begin{equation}
\mu = \lambda _{\max}\left(\sum_q \hat H_q^{(A)} \otimes \hat
H_q^{(B)}\right) -\lambda_{\min}\left(\sum_q \hat H_q^{(A)} \otimes \hat
H_q^{(B)}\right)  \notag
\end{equation}
is a constant that scales linear with the size $N$ of the surface separating
the subsystems.

In order to solve the differential inequality (\ref{final0}) we need an
expression or at least an estimate for $\mathrm{tr}\rho _{A}^{3}$ in terms
of the purity. With the help of Hardy's inequality $\left(
\sum_{k}a_{k}^{m}\right) ^{1/m}\leq \left( \sum_{k}a_{k}^{n}\right) ^{1/n}$
for any $a_{k}\geq 0$, and $m>n>0$, (see \cite{Hardy1952}), one can show
that $\mathrm{tr}\rho _{A}^{3}\leq \left( \mathrm{tr}\rho _{A}^{2}\right)
^{3/2}$. With this we find the following differential inequality for the
purity
\begin{equation}
\left\vert \frac{\mathrm{d}}{\mathrm{d}t}\mathrm{tr}\rho _{A}^{2}\right\vert
\leq 2\mu \sqrt{\left( \mathrm{tr}\rho _{A}^{2}\right) ^{3/2}-\left( \mathrm{%
tr}\rho _{A}^{2}\right) ^{2}}.  \label{eq:diffinequality}
\end{equation}%
In general the short time behavior is local, and eq. (\ref{eq:diffinequality})
partially supports this intuition, since the control parameter is $\mu$, the
spread of the local Hamiltonian. However, due to the second factor the dynamics
of the purity does not only depend on the local Hamiltonian, but also on the purity of the
initial state (see examples).

We can divide the left- and right-hand side by the square root term, which
after integration yields
\begin{equation}
\left\vert \frac{\mathrm{d}}{\mathrm{d}t}\arcsin \left( \left( \mathrm{tr}%
\rho _{A}^{2}\right) ^{1/4}\right) \right\vert \leq \frac{\mu }{2}.
\end{equation}%
Using
\begin{eqnarray}
\arcsin \left[ \left( \mathrm{tr}\rho _{A}^{2}\right) ^{1/4}\right]
&=&\int\limits_{0}^{t}\frac{\mathrm{d}}{\mathrm{d}\tau }\Bigl[\arcsin \left(
\left( \mathrm{tr}\rho _{A}^{2}\right) ^{1/4}\right) \Bigr]\mathrm{d}\tau +
\notag \\
&&+\arcsin \left( \left( \mathrm{tr}\rho _{A}^{2}\left( 0\right) \right)
^{1/4}\right)
\end{eqnarray}%
gives a solution of eq.(\ref{eq:diffinequality}) with the initial purity $%
\mathrm{tr}\rho _{A}^{2}\left( 0\right) $
\begin{eqnarray}
&&\sin ^{4}\left[ \max \left( -\frac{\mu }{2}t+\arcsin \left( \mathrm{tr}%
\rho _{A}^{2}\left( 0\right) \right) ^{1/4},0\right) \right] \leq \mathrm{tr}%
\rho _{A}^{2}  \notag \\
&&\enspace\leq \sin ^{4}\left[ \min \left( \frac{\mu }{2}t+\arcsin \left(
\mathrm{tr}\rho _{A}^{2}\left( 0\right) \right) ^{1/4},\frac{\pi }{2}\right) %
\right] .  \label{eq:arbitInitialCondition}
\end{eqnarray}%
If $\mathrm{tr}\rho _{A}^{2}\left( 0\right) =1$, i.e. if the subsystems are
uncorrelated in the beginning, the upper bound is trivial. The lower bound
reduces to
\begin{equation}
\mathrm{tr}\rho _{A}^{2}\geq \cos ^{4}\frac{\mu t}{2},\qquad \text{for}%
\enspace\mu t\leq \pi .  \label{eq:final}
\end{equation}%
We note that the lower bound becomes zero at $\mu t>\pi $ i.e. it reduces to
the trivial one.

The lower bound can be slightly improved, if we know the maximum Schmidt
rank $L_{\mathrm{max}}$ of the bipartite decomposition that can arise along
the evolution. One finds
\begin{equation}
\mathrm{tr}\rho_{A}^{2}\geq\cos^{4}\frac{\mu t}{2} + \frac{1}{L_{\mathrm{max}%
}-1}\sin^4\frac{\mu t}{2}.  \label{eq:final-2}
\end{equation}
Evidently $\mathrm{tr}\rho_A^2 \ge L_{\mathrm{max}}^{-1}$ as it should be.
Typically $L_{\mathrm{max}}\gg 1$ and thus the second term in eq.(\ref%
{eq:final-2}) is small. In the general case, $L_{\mathrm{max}}$ can be the
dimension of the smaller Hilbert space. So one must assume this, if there is
no better bound known a priori. In certain special cases as for the 1D
quantum Ising model with an initial state that factorizes in all sites, $L_{%
\mathrm{max}}=2$, and the lower bound (\ref{eq:final-2}) becomes exact, see
below.

The estimation (\ref{eq:final-2}) follows from the inequality
\begin{widetext}
\begin{equation}
tr\rho^{3}\leq\frac{1}{L_{\rm max}^{3}}\left[  \left(  1+\sqrt{\left(  L_{\rm max}-1\right)
\left(  L_{\rm max}\cdot tr\rho^{2}-1\right)  }\right)  ^{3}+\frac{\left(
L_{\rm max}-1-\sqrt{\left(  L_{\rm max}-1\right)  \left(  L_{\rm max}\cdot tr\rho^{2}-1\right)  }\right)
^{3}}{\left(  L_{\rm max}-1\right)  ^{2}}\right] \label{Inequality}%
\end{equation}
\end{widetext}which can be proven by the method of Lagrange multipliers.

Eqs. (\ref{eq:arbitInitialCondition}) and (\ref{eq:final}), resp. (\ref%
{eq:final-2}), provide an estimate for the purity for short times. As
expected from general quantum mechanical arguments the lower bound of the
purity decreases quadratically in time following $\sim -(t/\tau)^2$. The
characteristic time $\tau$ that defines the range of validity of the
quadratic time scaling is inverse proportional to the eigenvalue spread $\mu$
and scales linearly with the size of the surface $N$ separating the
subsystems and thus has an area-law behavior. In order to test the scaling
of the lower bound with the boundary size of the system we consider next a
pair of linear spin chains subject to an Ising interaction
\begin{eqnarray}
\hat H = \sum_{j=1}^N \hat \sigma_j^x \otimes \hat \tau_j^x,
\label{eq:isingsigmatau}
\end{eqnarray}
where $\hat \sigma$ and $\hat \tau$ are the Pauli operators of the two
chains respectively. If the initial state is
\begin{equation}
\left\vert \Downarrow_{A}\Uparrow_{B}\right\rangle =\left\vert \downarrow
\downarrow\dots\downarrow\right\rangle _{A}\otimes\left\vert \uparrow
\uparrow\dots\uparrow\right\rangle _{B}  \label{product}
\end{equation}
i.e. where all spins in chain $A$ are in the eigenstate of $\hat\sigma^z$,
resp. $\hat\tau^z$, with spin down and all spins in chain $B$ in the
corresponding spin up state, the purity is given by
\begin{equation}
\text{tr}\rho^{2}=\left( \cos^{4}t+\sin^{4}t\right) ^{N}
\label{noninteracting}
\end{equation}
This can be seen easily, as for the initial product state the total purity
can be represented as
\begin{equation}
\text{tr}\rho^{2}=\text{tr} {\prod\limits_{i=1}^{N}} \rho_{i}^{2}=\left(
\text{tr}\rho_{1}^{2}\right) ^{N}=\left( \cos^{4}t+\sin^{4}t\right) ^{N}
\label{totalpurity}
\end{equation}
because each spin pair of the coupled chains has a purity equal to $%
\cos^{4}t+\sin^{4}t$. We thus see, that the characteristic time $\tau$
scales as $\sqrt{N}$ for this choice of the initial state. However, if one
considers a Greenberger-Horne-Zeilinger type initial state
\begin{widetext}
\begin{equation}
\left\vert {\rm GHZ}\right\rangle_{A}\otimes\left\vert{\rm GHZ}\right\rangle_{B} =\left(  \frac{1}{\sqrt{2}%
}\left\vert ++\dots+\right\rangle _{A}+\frac{1}{\sqrt{2}}\left\vert
--\dots-\right\rangle _{A}\right)  \otimes\left(  \frac{1}{\sqrt{2}}\left\vert
++\dots+\right\rangle _{B}+\frac{1}{\sqrt{2}}\left\vert --\dots-\right\rangle
_{B}\right)  \label{GHX}%
\end{equation}
\end{widetext}
with $|\pm\rangle$ denoting eigenstates of $\hat\sigma^x$, resp. $\hat\tau^x$%
, one finds
\begin{equation}
\text{tr}\rho^{2}=\cos^{4}Nt+\sin^{4}Nt,  \label{purityEntangl}
\end{equation}
as can be shown by simple algebraic calculations. We see that (\ref%
{purityEntangl}) coincides with our estimation (\ref{eq:final-2}) since $%
\mu=2N$ and $L_{\mathrm{max}}=2$. In other words, our estimate is a tight
lower bound for the purity. In this case the system size scaling of the
characteristic time is $N$.

Note, that in the special case of an Ising Hamiltonian, our two-chains
example in fact represents also higher dimensional Ising lattices of
arbitrary size. Since all summands in the Ising Hamiltonian, say on a
hyper-cubic lattice commute, they can be absorbed in the quantum states of
the subsystems $A$ an $B$. They play no role for the entanglement, which is
only created by the terms directly coupling $A$ and $B$. But those can
always be written in the form (\ref{eq:isingsigmatau}), no matter what the
spacial dimension of the surface is.

For short times, the quadratic estimate (\ref{eq:final}) is much better than
any exponential one (\ref{eq:largelimeestimation}) (see below). Furthermore
it is of fundamental importance. It shows e.g. that a sequence of frequent
projections of the system onto non-entangled states (i.e. no entanglement
within the system) at a rate larger than $\mu$ will prevent the build-up of
such an entanglement in full analogy to the quantum Zeno effect \cite%
{Facchio2001a}.


\section*{III -- Long time behavior}


The long time behavior of the purity can be obtained from the known upper
linear-in-time bound of the entropy \cite{Bravyi2006,Eisert2006,Lieb1972}.
Indeed, by using the convexity of $-\ln x,$ we immediately get
\begin{equation}
S=-\sum\limits_{i} \xi_{i}\ln\xi_{i}\geq-\ln \sum\limits_{i}
\xi_{i}^{2}=-\ln \mathrm{tr}\rho_{A}^{2},  \label{eq:entropy-bound}
\end{equation}
We thus have
\begin{equation*}
\mathrm{tr}\rho_{A}^{2}\geq\exp(-S)\geq\exp\left( -c_{0}\right)\, \exp\left(
-c_{1}t\right) ,
\end{equation*}
where $S\leq c_{0}+c_{1}t$, $c_{0}$ and $c_{1}$ being positive constants.

In the following we show that (\ref{eq:entropy-bound}) can also be obtained
in a simple way from our approach without the necessity to invoke the rather
involved proof of the linear-in-time-bound of the entropy.

In order to find a suitable estimate for the long-time behavior of the
purity one has to find a different way to bound the right hand side of eq. (%
\ref{eq:purity}). The interference effects become negligible at $t\gg\frac{1%
}{\mu}$ and therefore we may use inequalities of the Schwartz type. In other
words, we may sum all interactions (matrix elements) in modulus instead of
amplitude. In this case one finds
\begin{eqnarray}
&&\left\vert \frac{\mathrm{d}}{\mathrm{d}t} \mathrm{tr}\rho_{A}^{2}\right%
\vert \leq 2\sum\limits_{q,\alpha,\beta} \sqrt{\xi_{\alpha}\xi_{\beta}}%
\left\vert \xi_{\alpha}-\xi _{\beta}\right\vert \times\,  \notag \\
&&\enspace\times \left\vert \left\langle \phi_{\alpha}^{\left( A\right)
}\right\vert \hat H_{q}^{\left( A\right) }\left\vert \phi _{\beta}^{\left(
A\right) }\right\rangle \left\langle \phi _{\alpha}^{\left( B\right)
}\right\vert \hat H_{q}^{\left( B\right) }\left\vert \phi_{\beta}^{\left(
B\right) }\right\rangle \right\vert \\
&& \leq \sqrt{2}\sum\limits_{q,\alpha,\beta} \xi_{\alpha}^{2}\left\vert
\left\langle \phi_{\alpha}^{\left( A\right) }\right\vert \hat H_{q}^{\left(
A\right) }\left\vert \phi_{\beta }^{\left( A\right) }\right\rangle
\left\langle \phi_{\alpha}^{\left( B\right) }\right\vert \hat H_{q}^{\left(
B\right) }\left\vert \phi _{\beta}^{\left( B\right) }\right\rangle
\right\vert .  \notag
\end{eqnarray}
Here we have used
\begin{eqnarray*}
\sqrt{2\xi_{\alpha}\xi_{\beta}}\left\vert \xi_{\alpha}-\xi
_{\beta}\right\vert \leq\frac{\xi_{\alpha}^{2}+\xi_{\beta }^{2}}{2}.
\end{eqnarray*}
Making use of Schwartz's inequality, we obtain
\begin{eqnarray*}
&&\sum\limits_{\beta} \left\vert \left\langle \phi_{\alpha}^{\left( A\right)
}\right\vert \hat H_{q}^{\left( A\right) }\left\vert \phi_{\beta}^{\left(
A\right) }\right\rangle \left\langle \phi_{\alpha}^{\left( B\right)
}\right\vert \hat H_{q}^{\left( B\right) }\left\vert \phi_{\beta}^{\left(
B\right) }\right\rangle \right\vert \leq  \notag \\
&& \sqrt{\sum\limits_{\beta} \left\vert \left\langle \phi_{\alpha}^{\left(
A\right) }\right\vert \hat H_{q}^{\left( A\right) }\left\vert
\phi_{\beta}^{\left( A\right) }\right\rangle \right\vert ^{2}
\sum\limits_{\beta} \left\vert \left\langle \phi_{\alpha}^{\left( B\right)
}\right\vert \hat H_{q}^{\left( B\right) }\left\vert \phi_{\beta}^{\left(
B\right) }\right\rangle \right\vert ^{2}}  \notag \\
&&= \sqrt{\left\langle \phi_{\alpha}^{\left( A\right) }\right\vert \left(
\hat H_{q}^{\left( A\right) }\right) ^{2}\left\vert \phi_{\alpha}^{\left(
A\right) }\right\rangle \left\langle \phi_{\alpha}^{\left( B\right)
}\right\vert \left( \hat H_{q}^{\left( B\right) }\right) ^{2}\left\vert
\phi_{\alpha}^{\left( B\right) }\right\rangle } \\
&&\leq \sqrt{\lambda_{\max}\left[ \left( \hat H_{q}^{\left( A\right)
}\right) ^{2}\right] \lambda_{\max}\left[ \left( \hat H_{q}^{\left( B\right)
}\right) ^{2}\right] }.  \notag
\end{eqnarray*}
We thus arrive at
\begin{equation}
\left\vert \frac{\mathrm{d}}{\mathrm{d}t}\mathrm{tr}\rho_{A}^{2}\right\vert
\leq\chi \, \mathrm{tr}\rho_{A}^{2},  \label{eq:purityLongtime}
\end{equation}
where
\begin{equation*}
\chi=\sqrt{2}\, \sum\limits_{q} \sqrt{\lambda_{\max}\left[ \left( \hat
H_{q}^{\left( A\right) }\right) ^{2}\right] \lambda_{\max}\left[ \left( \hat
H_{q}^{\left( B\right) }\right) ^{2}\right] }.
\end{equation*}
The solution of this differential inequality with the initial condition $%
\mathrm{tr}\rho_{A}^{2}\left( 0\right) =1$ is
\begin{equation}
\mathrm{tr}\rho_{A}^{2}\geq\exp\Bigl( -\chi t\Bigr) .
\label{eq:largelimeestimation}
\end{equation}


\section*{IV -- Numerical examples: spin--$\frac{1}{2}$ XX- and XXZ-chains,
multidimensional quantum Ising model}

\label{section3}


\begin{figure}[ptb]
\begin{center}
\includegraphics[width=1.0\columnwidth]{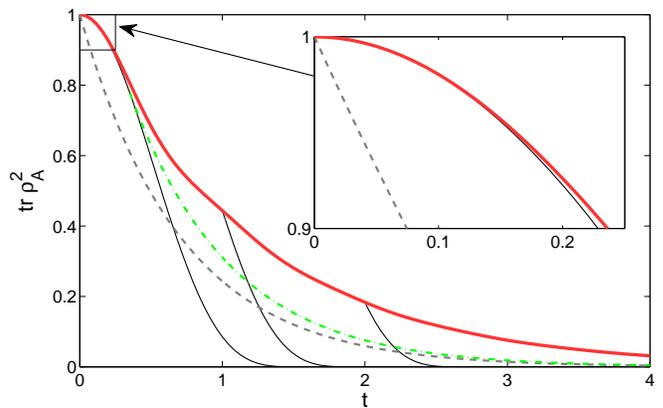}
\end{center}
\caption{(color online) Time evolution of the purity in the $80+80$ site XX
chain compared to the bounds (\protect\ref{eq:final}) and (\protect\ref%
{eq:largelimeestimation}), see text. The inset shows a closeup for short
times. Note that $t$ has no units since we have choosen the dimensionless Hamiltonian (\ref{eq:def_h_xx}).}
\label{Fig}
\end{figure}

To illustrate the quality of the bounds given in (\ref{eq:final}) and (\ref%
{eq:largelimeestimation}) we perform exact numerical simulations for a large
1D spin system as well as for two coupled spin chains of small size. We
first consider the spin-$\frac{1}{2}$ XX--model
\begin{equation}
\hat H=-\frac{1}{2}\sum_{j}\left(\hat\sigma_{j}^{x}\hat\sigma_{j+1}^{x}+\hat%
\sigma_{j}^{y}\hat\sigma_{j+1}^{y}\right) .
\label{eq:def_h_xx}
\end{equation}
For this model the constants which enter our estimates take on the values $%
\mu=2$ and $\chi=\sqrt{2}$. We look at a chain of $160$ spins and in order
to maximize the dimension of the subsystems we choose an equal partition
with $A$ ($B$) being the left (right) half chain. The initial state of the
system is taken to be a product state, specifically the anti-ferromagnetic
state
\begin{equation}
\left\vert \Psi_{A}(t=0)\right\rangle =\left\vert \Psi_{B}(t=0)\right\rangle
=\left\vert \uparrow\downarrow\uparrow\downarrow\uparrow\downarrow
\cdots\uparrow\downarrow\right\rangle.
\end{equation}
We choose this particular initial state in order to have a large maximum
entropy since
$|\Psi_A(t=0)\rangle$ corresponds to half filling, so the dimension of the
Hilbert space accessible with respect to the present conservation of the
total z--magnetization is maximized. The purity is initially $1$. Fig. \ref%
{Fig} shows the evolution of the purity over time. The red, straight, thick
line is the numerical results from our simulation using the time-evolving
block decimation (TEBD) method \cite{Vidal2003,Vidal2004}. This results can
be considered numerically exact as discussed below. The solid, black,
thinner lines show the bounds (\ref{eq:final}). The one starting at $t=0$
indicates that our bound is optimal up to second order for times short
compared to the inverse coupling between $A$ and $B$, if we start initially
from a pure state. However when starting from an initially entangled state,
we can only expect to get agreement up to zeroth order from (\ref%
{eq:arbitInitialCondition}), as illustrated by the black, solid, thinner
lines starting at $t=1$ and $t=2$. While the exponential lower bound (\ref%
{eq:largelimeestimation}), plotted in the dashed line, is bad for short
times, it has the property of remaining finite for all times in contrast to (%
\ref{eq:final}). So one can smoothly concatenate the two bounds at time
\begin{equation}
t_{1}=\frac{2}{\mu}\arctan\left( \frac{\chi}{2\mu}\right) ,
\end{equation}
assuming we started with a pure state at $t=0$. This combined bound is
superior to both the quadratic short-time and exponential long-time
estimates and is shown as a green, dot-dashed line.

\begin{figure}[ptb]
\begin{center}
\includegraphics[width=1.0\columnwidth]{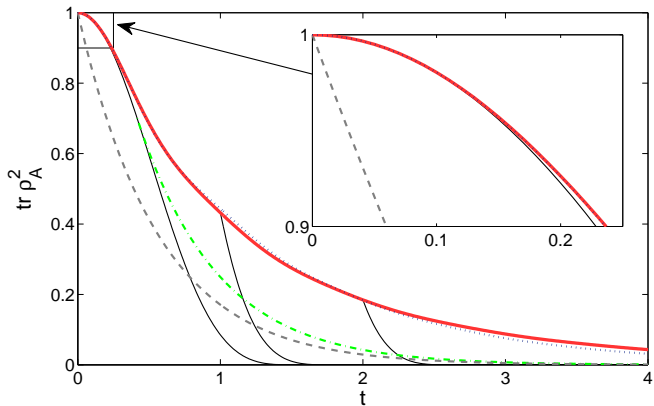}
\end{center}
\caption{(color online) Time evolution of the purity in the $80+80$ site XXZ
chain for $\Delta=\frac12$ compared to the bounds (\protect\ref{eq:final})
and (\protect\ref{eq:largelimeestimation}) in analogy to Fig. \protect\ref%
{Fig}. The inset shows a closeup for short times. The dotted lines show the
purity in the XX--model (Fig. \protect\ref{Fig}) for comparison.}
\label{Fig2}
\end{figure}

Analogous calculations where also done for the spin-$\frac{1}{2}$ XXZ--model
\begin{equation}
\hat H=-\frac{1}{2}\sum_{j}\left( \hat\sigma_{j}^{x}\hat\sigma_{j+1}^{x}
+\hat\sigma_{j}^{y}\hat\sigma_{j+1}^{y}
+\Delta\hat\sigma_{j}^{z}\hat\sigma_{j+1}^{z}\right),
\end{equation}
choosing the anisotropy $\Delta$ to be $\frac12$. This again yields a
constant of $\mu=2$ for the short time behavior. But while the exponential
bound increases to $\chi=\frac{5}{2\sqrt{2}}$ and one could expect a faster
decay of the purity due to the fact that this system can not be mapped to
free fermions, the true curves are very much alike for both systems, see
Fig. \ref{Fig2}.

The numerical calculation was done using a matrix dimension of $D=500$, a
time-step width of $0.01$ in a fourth order Trotter decomposition, and
exploiting the conservation of the total magnetization explicitly. Although
the dimension of the subsystems is adaptively truncated to $D$, this cannot
introduce error on the timescale plotted, since the purity remains well
above the minimal value of $1/D$ representable using matrix product states.
Thus all relevant states are included by the algorithm.

\begin{figure}[ptb]
\begin{center}
\includegraphics[width=1.0\columnwidth]{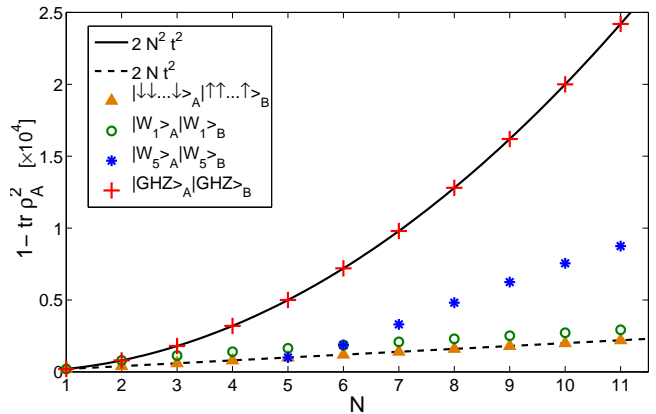}
\end{center}
\caption{(color online) Short time scaling of the purity with boundary size
for a pair of coupled spin chains with Ising interaction. Exact
diagonalization was done for $t=0.001$, such that the quadratic order in (%
\protect\ref{eq:final}) and (\protect\ref{totalpurity}) suffices to describe
the result. While the GHZ-like initial states show the fastest increase of
entanglement ($\protect\tau\sim N$, eq. (\protect\ref{eq:final}), straight
line), and the product like ones stick to the moderate $\protect\tau\sim%
\protect\sqrt{N}$ (eq. (\protect\ref{totalpurity}), dashed line), W-type
states show an intermediate behavior. }
\label{fig:ising}
\end{figure}

Figure \ref{fig:ising} illustrates the tightness of (\ref{eq:final}) for
proper initial conditions. It shows $1-\mathrm{tr}\rho_{A}^2$ in a system of
two Ising spin chains of length $N$, subject to the Hamiltonian (\ref%
{eq:isingsigmatau}), after a short time of $t=0.001$ of evolution.
Ising-type couplings inside the chains where also taken into account, but do
not contribute to the entanglement between the two chains. While for simple
product states between the sites, one can expect a scaling of $\tau\sim\sqrt{%
N}$ (dashed line), we know from (\ref{purityEntangl}), that there are in
fact initial states, that contain sufficient entanglement along the
boundary, to give$\tau\sim{N}$, i. e., where (\ref{eq:final}) is tight. The
different symbols correspond to different initial states, the evolution of
which was calculated via an exact diagonalization. As already seen above, we
have a $N$ scaling for an initial product state like (\ref{product}), while
we get an exact $N^2$ scaling for GHZ-like initial states (\ref{GHX}). Also
shown are other states like the W-type states $\left\vert
W_p\right\rangle_A\otimes\left\vert W_p\right\rangle_B$, where
\begin{equation}
\left\vert W_p\right\rangle = \sum_{1\le j_1<j_2<\dots<j_p\le N}
\hat\sigma^x_{j_1}\hat\sigma^x_{j_2}\dots\hat\sigma^x_{j_p} \left\vert
\downarrow\downarrow\dots\downarrow\right\rangle.  \label{eq:Wp}
\end{equation}

\section*{Summary}


In summary we derived an upper bound for the time evolution of the bipartite
entanglement in quantum lattice models in terms of a lower bound to the
subsystem purity. As one would expect from general quantum mechanical
considerations the purity decreases for short times quadratically in time.
The corresponding characteristic time was shown to be limited by the spread
of the eigenvalues of the part of the Hamiltonian that accounts for the
interaction between the partitions. The latter scales linear with the size
of the surface separating the two partitions and thus the entanglement
follows an area-law behavior. For larger times we derived a lower bound of
the purity that decrease exponentially in time. The latter is equivalent to
the known linear increase of the entanglement entropy in the long-time
limit. The existence of a quadratic short-time bound means that a
sufficiently frequent sequence of projections to non-entangled states, as
for example due to a dissipative process, can prevent the build-up of
entanglement within the lattice system.


\section*{Acknowledgment}

The authors thank D. Bruss and J. Anglin for fruitful discussions. This work
has been supported by the DFG through the SFB-TR49 and by the graduate
school of excellence MATCOR.


\end{document}